\documentclass[namedreferences]{kluwer}
\usepackage[dvips]{graphicx}

\begin{document}
\begin{article}
\begin{opening}
\title{HIGH PERFORMANCE CORRUGATED FEED HORNS \\
FOR SPACE APPLICATIONS
AT MILLIMETRE \\
WAVELENGTHS}
%\thanks{This work has been supported by the Italian Space Agency (ASI), bla bla bla}}
\author{F. \surname{Villa}}
\author{M. \surname{Sandri}}
\author{N. \surname{Mandolesi}}
\institute{IASF--CNR, Sezione di Bologna, Italy,}
\author{R. \surname{Nesti}}
\institute{INAF -- Osservatorio Astrofisico di Arcetri, Firenze, Italy}
\author{M. \surname{Bersanelli}}
\institute{Universit\`a degli Studi di Milano, Milano, Italy}
\author{A. \surname{Simonetto}}
\author{C. \surname{Sozzi}}
\author{O. \surname{D'Arcangelo}}
\author{V. \surname{Muzzini}}
\institute{Istituto di Fisica del Plasma, CNR, Milano, Italy}
\author{A. \surname{Mennella}}
\institute{IASF--CNR, Sezione di Milano, Italy}
\author{P. \surname{Guzzi}}
\author{P. \surname{Radaelli}}
\author{R. \surname{Fusi}}
\author{E. \surname{Alippi}}
\institute{LABEN, Vimodrone, Milano, Italy}
\runningtitle{CORRUGATED FEED HORNS FOR SPACE APPLICATIONS}
\runningauthor{F. VILLA ET AL.}

\begin{ao}\\
Fabrizio Villa \\
Istituto IASF/CNR Sezione di Bologna\\
Via P. Gobetti, 101 \\
I-40129, Bologna, Italy
\end{ao}


\begin{abstract}
We report on the design, fabrication and testing of a set of high
performance corrugated feed horns at 30 GHz, 70 GHz and 100 GHz, built as
advanced prototypes for the Low Frequency Instrument (LFI) of the ESA {\sc
Planck} mission. The electromagnetic designs include linear (100 GHz)  and
dual shaped (30 and 70 GHz) profiles. Fabrication has been achieved by
direct machining at 30 GHz, and by electro-formation at higher
frequencies. The measured performances on side lobes and return loss meet 
the stringent {\sc Planck} requirements over the large (20\%)  instrument bandwidth. 
Moreover, the advantage in terms of main lobe shape and side lobes levels of the dual profiled designs has been demonstrated.
\end{abstract}
\end{opening}

%%%%%%%%%%%%%%%%%%%%%%
\section{Introduction}
%%%%%%%%%%%%%%%%%%%%%%

Corrugated horn antennas exhibit a combination of highly desirable characteristics, such as 
high beam symmetry, low cross polarization, large bandwidth, low level of side lobes,
good return loss and low attenuation \cite{Olver92, Olver94}. 
In the past two decades corrugated horns have been extensively used,
either alone or coupled to reflectors, in a number of ground based and space astrophysical observations in the microwave and millimeter wavelength range.
In particular, high performance designs have been developed for several 
experiments devoted to the study of the Cosmic Microwave Background (CMB), including the NASA space missions COBE/DMR \cite{Toral89}, and WMAP\footnote{\tt http://map.gsfc.nasa.gov/}\cite{Bennett03, Barnes02}.

Corrugated feed horns will be also implemented in the two instruments
(LFI and HFI) onboard the ESA {\sc Planck} satellite\footnote{\tt http://astro.estec.esa.nl/SA-general/Projects/Planck/}\cite{Villa03}.
Both LFI, the Low Frequency Instrument
\cite{Mandolesi98}, and HFI, the High Frequency Instrument
\cite{Puget98}, share the focal region of a 1.5 meter off-axis
dual reflector aplanatic optimized telescope \cite{Dubruel00, Villa01}. 
Together, they give {\sc Planck} unprecedented sensitivity and
angular resolution over the full spectral range between 30 GHz and
857 GHz, ensuring the imaging power, the redundancy, and the control
of systematic effects and of foreground emissions needed to
achieve the mission science goals. Both {\sc Planck} instrument teams 
selected dual
profiled conical corrugated horns as the best
option in the range $30-350$ GHz since they guarantee 
superior performance compared with linear conical horns. 

Specifically, the dual profiled design permits to optimise the radiation characteristics
by reaching a nearly perfect Gaussian shape of the main lobe and level of sidelobes well below $-30$ dB. While in linear horns the pattern is controlled by the diameter and the flare angle, in dual profiled horns five parameters can be optimised, leading to significantly improved characteristics both from the electromagnetic and mechanical point of view. 

The LFI is composed by an array of 12 feed horns at 100 GHz, 6 at 70 GHz, 3
at 44 GHz, and 2 at 30 GHz, feeding 46 radiometers each
producing two output channels for a total of 94 detectors.
Several prototype feed horns have been produced since
1998 from 25 GHz to 140 GHz by the LFI instrument team
\cite{Bersanelli98, Guzzi99, Villa97, Villa98}.
A set of three corrugated feed horns have been designed, developed, and
tested as advanced prototypes for the LFI feeds, with 
the purpose of demonstrating the maturity of the design with respect to the
flight performance requirements, including the fabrication
technology. 
One feed at $30$ GHz has been developed according to the Flight Model mechanical and electromagnetic specifications. One feed at $70$ GHz has been developed to be representative both mechanically and electromagnetically of the six Flight Models foreseen. One 100 GHz feed has been developed specifically to investigate the manufacturing capability to reproduce the corrugation detail. For this feed a linear design with $90$ corrugations has been chosen and no mechanical requirements have been imposed.

According to the development philosophy of the LFI, these prototypes constitute the Elegant
Bread Board (EBB) model. 

\section{Electromagnetic and mechanical requirements}

The baseline design for all the 23 LFI horns foresees a profiled shape of the corrugated
section. Dual profiled corrugated horns have been selected for their compactness both 
in length and diameter and their superior performances in terms of ability to control
the shape of the main lobe, the level of the side lobes and the phase center location. 
In the linear corrugated horn only the hybrid $HE_{11}$ mode propagates. Typically this mode is selected by having a mixture of $TE_{11}$ and $TM_{11}$ that are running in phase and with the same propagation constant inside the horn. In dual profiled horns, other modes are excited in a controlled way to modify the field distribution at the aperture, being the wanted distribution to be as good as gaussian. In this way, the main lobe exhibits a Gaussian shape and sidelobes may be reduced because of the low level of the field at the aperture rim.

\begin{figure}[!ht]
\centerline{\includegraphics[width=28pc]{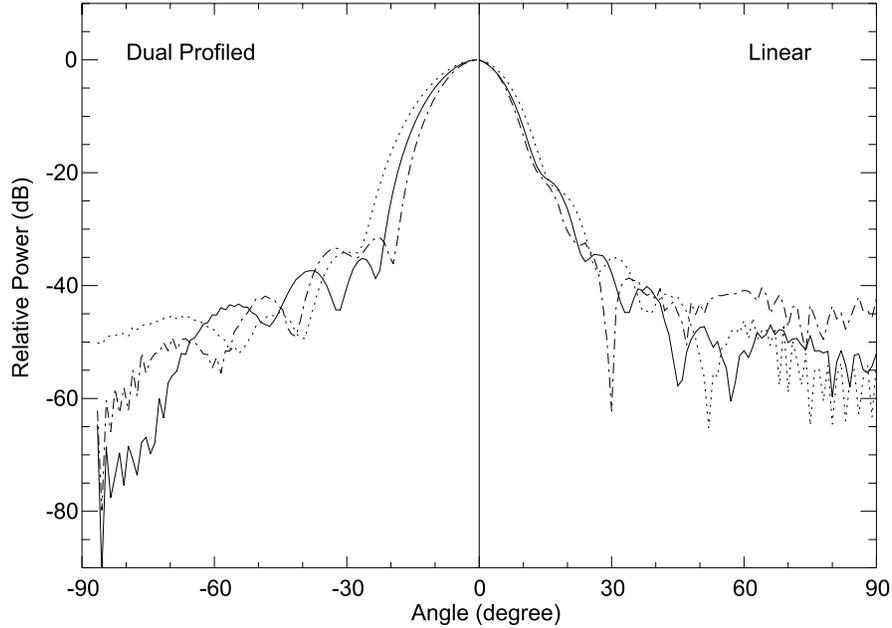}}
\caption{E plane measurements at 63 GHz (dashed line), 70 GHz (solid line) and 77 GHz (dash--dotted line) 
of the V--band dual profiled horn (left side) and E plane measurements at 90 GHz 
(dashed line), 100 GHz (solid line) 
and 110 GHz (dash--dotted line) of the W--band linear horn (right side).
The larger spread in frequency of the dual profiled design is evident.} 
\label{fig:linear_dp}
\end{figure}

On the other hand, 
the amplitude pattern of this kind of design exhibits a more pronounced variation in 
frequency compared to the linear corrugated horns 
(see Figure \ref{fig:linear_dp}).

Because of the complexity of the LFI Focal Plane Unit layout, different designs are 
foreseen even at the same frequency. A total of twelve different designs are needed 
with the length, the diameter, and the location of the phase center varying with the position 
of the feeds in the focal plane. 

The three prototypes were developed using a different approach for each 
design. The main performances of the horns required by the Flight Models (FM) compared to the 
EBB models are summarized in Table \ref{tab:chars}. 

In order to satisfy the mechanical constraints imposed by the focal plane layout, both
the 30 and 70 GHz feeds need to be short 
without any degradation of the radiation pattern. These mechanical requirements 
have been maintained in the EBB design, which satisfy also the flight model electromagnetic specifications. 
The 100 GHz one is a linear corrugated feed horn with very thin grooves
on the corrugations near the throat (Table \ref{tab:ch}), mainly to investigate the manufacturing capability.

The baseline performances and characteristics of the LFI feed
horns are shown in Table \ref{tab:chars}. Edge Taper values have
been defined for each feed horn, depending on its location in the focal plane.  
LFI horns are located off--axis and each feed horn illuminates the telescope in a different way. As a consequence the edge taper definition may differ. This is the case of the $70$ and $100$ horns. The edge taper definition of these horns varies and in table \ref{tab:chars} only the minimum and maximum edge taper values are reported, for simplicity.

\begin{table}[hptb]
\caption{Requirements of all the LFI baseline (FM) feed horns (first four
lines) and characteristics of the prototypes (EBB).
ET is the Edge Taper, SLL is the maximum Side Lobes Level, X-pol
is the cross polarization, RL is the return loss and IL is the
Insertion loss. For the 70 and 100 GHz horns the ET varies, depending on 
the position of the feed horn in the focal plane.}
\begin{tabular*}{\maxfloatwidth}{l l l l l l l}\hline
$\nu_0$  &  $\Delta \nu / \nu_0$  &  ET & SLL &X-pol
&RL & IL  \\
(GHz) &      &  (dB) &  (dB)& (dB)& (dB)& (dB) \\
\hline
\multicolumn{7}{c}{\sc LFI FM Corrugated Feed Horns}\\

\hline
30  & 20 \%  &  $30.0 ~@~ 22^\circ$        & $\leq -35$  & $\leq -30$ & $\leq -25$ & $\leq -0.10$  \\
44  & 20 \%  &  $30.0 ~@~ 22^\circ$        & $\leq -35$  & $\leq -30$ & $\leq -25$ & $\leq -0.10$  \\
70  & 20 \%  &  $20.5 - 25.0 ~@~ 22^\circ$ & $\leq -35$  & $\leq -30$ & $\leq -25$ & $\leq -0.10$  \\
100 & 20 \%  &  $21.0 - 23.3 ~@~ 22^\circ$ & $\leq -35$  & $\leq -30$ & $\leq -25$ & $\leq -0.10$  \\
\hline
\multicolumn{7}{c}{\sc LFI EBB Corrugated Feed Horns}\\
\hline
30  & 20 \%  &  $30.0 ~@~ 22^\circ$        & $\leq -35$  & $\leq -30$ & $\leq -25$ &  \\
70  & 20 \%  &  $21.6 ~@~ 22^\circ$      & $\leq -35$  & $\leq -30$ & $\leq -25$ &   \\
100 & 20 \%  &  $30.0 ~@~ 22^\circ$        & $\leq -35$  & $\leq -30$ & $\leq -25$ &  \\
\hline
\end{tabular*}
\label{tab:chars}
\end{table}

With these prototypes, the design, the development and the manufacturing processes have been optimised and completely characterized.

%%%%%%%%%%%%%%%%%%%%%%%%%%%%%%%%
\section{Horn design}
%%%%%%%%%%%%%%%%%%%%%%%%%%%%%%%%

The design procedure can be summarized as follows. As a first step
the aperture diameter was chosen analytically to give the desired
co-polar pattern beamwidth. For the linear design, the diameter has been evaluated by means of simulations carried out by calculating the radiation pattern under the pure $HE_{11}$ mode propagation in circular waveguide with a phase distribution given by a spherical phase front with centre in the apex of the horn \cite{Sletten88}. This procedure 
permitted to perform several simulations with different diameters and flare angles in a reasonable time.
Since the depth and shape of the corrugations determine the cross-polar radiation
characteristics, their geometry was selected to give the
minimum level of cross polarization at the center frequency. The corrugations 
are about a quarter of a wavelength deep and there are
at least three corrugations per wavelength in order to well approximate a
continuous impedance surface. The junction and the throat region were designed to optimize the impedance match to the smooth wall waveguide. For this purpose the first slot seen by the
smooth wall waveguide is approximately half a wavelength deep with the following few slots (about five) constituting a transition region to the quarter wavelength - steady state depth of the remaining part. Also the teeth and groove length need to be varied continuously between the
first corrugations and the steady-state region: a good input matching
requires thick teeth and thin grooves, which are used at the beginning,
while, on the contrary, HE11 propagation is well supported by thick grooves
and thin teeth. Having this in mind, the throat geometry has been obtained. 
In the case of dual profiled designs, an optimization routine has been applied. Six 
optimization parameters has been used. They are the slot depth, the teeth and groove width of the first and the steady-state corrugation. These parameters have been varied under a linear low and assuming a definite number of corrugations.
Finally, the flare angle and the diameter were optimised by attempts, taking into account the length of the horn and the co-polar pattern characteristics.
For this optimisation, simulations have been performed considering the detail of the corrugations.
The flare angle does not have to be constant along the horn: a profiled horn is a horn with
a flare angle which changes along the horn with a sine square
profile. If it is used as a feed for a reflector, as in our case, it can 
produce important improvements like higher gain and efficiency, very low
level of cross polarization, low mass and compactness.

\begin{figure}
\centerline{\includegraphics[width=26pc]{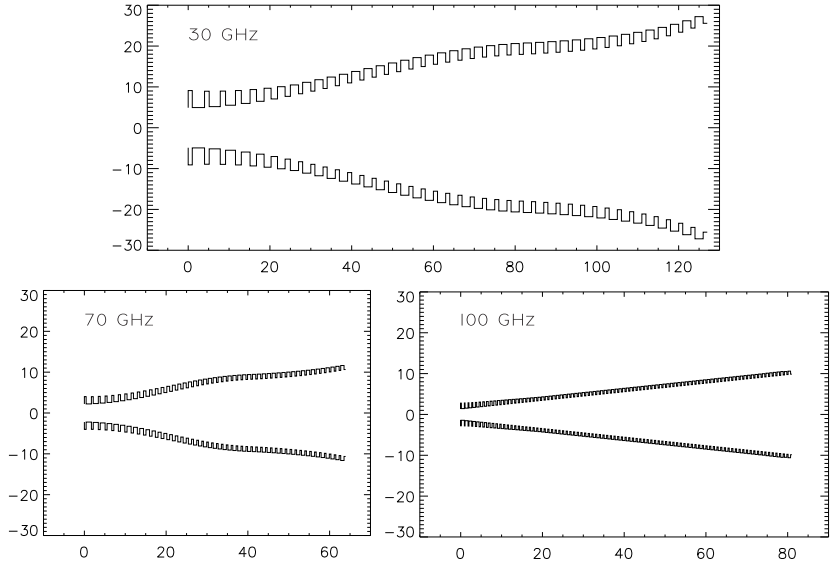}}
\caption{Corrugation profiles of the three horns (same scale for the three panels). 
Dimensions are in mm. Note the compactness of the 70 GHz horn
which is shorter than the 100 GHz one.}
\label{fig:profilo}
\end{figure}

Figure \ref{fig:profilo} reports the corrugation profile of the three designs.
In terms of wavelength normalized total length ($L_\nu$, where $\nu$ is the frequency in GHz), the 30 GHz horn is the shortest one, with
$L_{30} = 12.7$. For the horn at $70$ GHz, the length is $L_{70} = 14.91$. At 100 GHz, 
the linear horn is $L_{100} = 27$ long. The advantage of the dual profiled design can be appreciated by these numbers: the $100$ GHz horn results a factor of $2$ longer (in normalized length) than the $30$ GHz one; moreover, for a given edge taper (the 30 GHz and the 100 GHz horns exhibit the same edge taper), the 30 GHz horn results 30\% smaller in normalised diameter than the 100 GHz one. 

The main geometrical characteristics of the prototypes
are reported in Table \ref{tab:ch} and Figure \ref{fig:profilo}. 

\begin{table}[!ht]
\caption{Geometry of the EBB LFI corrugated horns. $L$: horn
length, $R_{th}$: throat radius, $R_{ap}$: aperture radius,
$d_{th}$: corrugation depth at the throat, $w_{th}$: corrugation
width at the throat, $d_{ap}$: corrugation depth at the aperture,
$w_{ap}$: corrugation width at the aperture. All dimensions are in
mm. $\alpha_{max}$: flare angle (degrees). For dual profiled horns
the flare angle is not indicated. The aperture diameters have been chosen for matching the Edge Taper 
requirements of the {\sc Planck} telescope. The inner region of each horn 
has been chosen in order to minimize the return loss.}
\label{tab:ch}
\begin{tabular*}{\maxfloatwidth}{c c c c c c c c c}\hline
Horn    & $L$    & $R_{th}$ & $R_{ap}$ & $d_{th}$ & $w_{th}$ & $d_{ap}$ & $w_{ap}$ & $\alpha_{max}$ \\
\hline
30 GHz  &$146.41$&  $4.90$  & $25.57$  & $4.2$    & $1.0$    & $2.80$   & $2.00$   &    $-$          \\
70 GHz  &$77.80$ &  $2.21$  & $11.766$ & $1.923$  & $0.206$  & $1.289$  & $0.90$   &    $-$          \\
100 GHz &$88.27$ & $1.337$  & $9.85$   & $1.453$  & $0.117$  & $0.886$  & 0.60     &    $6.0$        \\
\hline
\end{tabular*}
\end{table}

\subsection{Linear design}

The electromagnetic design of the 100 GHz horn has been obtained by simulations with Mode Matching 
\cite{Olver94}.
The flare angle has been defined at $6^\circ$ in order to maintain the phase error at the 
aperture at an acceptable level and to prevent pattern degradations.
Although the linear designs are well consolidated, this model represent a challenge from the  manufacturing point of view due to the corrugation dimensions. No mechanical constraints have been fixed on this prototype, while stringent requirement have been assumed on the return loss and beam symmetry. As suggested by Xiaolei (1993) an excellent return loss could be reached by gradually increasing the corrugations width near the throat section. This 
scheme has been applied at best by setting the maximum acceptable return loss at $-35$ dB over the whole $20\%$ of bandwidth.

\subsection{Dual Profiled Design}
The corrugation profile applied for the $30$ GHz and the $70$ GHz horns, is a mixture of a sine--squared section \cite{Olver88}, starting from the throat, and an exponential section near the aperture plane \cite{Teniente01, Gentili00}. 
The dual profiled design has been consolidated according to the corrugation profile, $R(z)$, 
\begin{eqnarray}
R(z) = &&R_{th} + (R_s - R_{th}) \left[ (1-A)\frac{z}{L_s} + A \sin^2
\left(\frac{\pi}{2}\frac{z}{L_s} \right) \right] \\
&&0  \leq z  \leq L_s \nonumber
\end{eqnarray}
\noindent
in the sine--squared section, and
\begin{eqnarray}
R(z) =&& R_s + e^{\alpha\left (z-L_s\right)}-1;\alpha = \frac{1}{L_e} ln(1+R_{ap}-R_s) \\
&&L_s \leq  z \leq L_e + L_s \nonumber
\end{eqnarray}
\noindent 
in the exponential region. Here, $R_{th}$ is the throat radius, $R_s$ is the sine squared region
end radius (or exponential region initial radius), $R_{ap}$ is the
aperture radius, $L_s$ is the sine squared region length and $L_e$
is the exponential region length. The parameter $A$ ($0 \le A \le 1$)
modulates the first region profile between linear and pure sine
squared type. The parameters $L_e/(L_e+L_s)$, $A$ and $R_s$ can be
used to control, as far as possible, the position and frequency
stability of the phase center and the compactness of the
structure.

The electromagnetic simulations have been carried out using a hybrid method
based on Mode Matching (MM) technique combined with the Moment Method (MoM)\cite{Coccioli96}. 
The MoM is used for the calculation of the radiation pattern taking into account the field 
distribution on the aperture plane and the currents traveling on the external part of the horn.

This hybrid technique is based on a particular formulation of the equivalence principle and the generalised scattering matrix approach which has the power to give directly the return loss at the throat with the contributions coming both from the internal corrugated region and the external feed shape. Moreover the use of the MoM makes it possible the
accurate calculation of the currents on the horn exterior conductor surface and the fields on the aperture, which are the sources of the electromagnetic field. From the sources, the beam pattern is easily computed by using the radiation integral. Finally an optimisation algorithm has been used to match the required electrical performances and mechanical properties by searching for the best horn geometry; edge taper, side lobe level, phase center and return
loss have been considered in the cost function to be minimised while the corrugation geometry and the parameters defining the corrugation profile have been used as optimisation variables.

%%%%%%%%%%%%%%%%%%%%%
\section{Fabrication}
%%%%%%%%%%%%%%%%%%%%%

\begin{figure}
\centerline{\includegraphics[width=28pc]{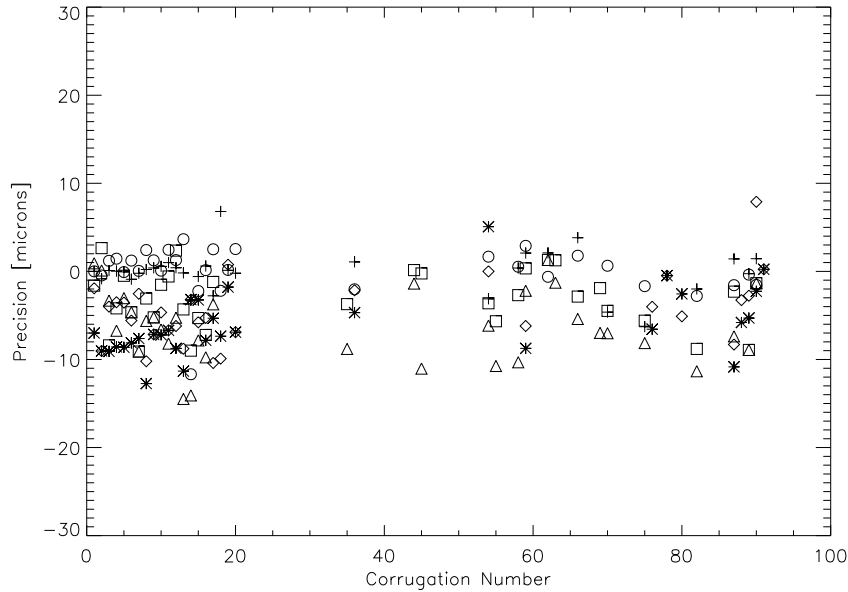}}
\caption{Differences ($\mu$m) between the theoretical dimensions and the  measured dimensions of two 100 GHz mandrels. On the abscissa, the corrugation number is
reported, from the throat to the aperture. Crosses and circles are referred to the deviation between the measured and theoretical dimensions, along the horn axis, of the grooves. Asterisks
(triangles) and diamonds (squares) are the differences of the groove (teeth) dimensions perpendicular to the horn axis.}
\label{fig:mandrino}
\end{figure}

Because of the small corrugation size, electroforming
has been chosen to manufacture the 70 and 100 GHz horns,
while 30 GHz horn has been built using electro-erosion. In
the first case, a pure aluminum master is directly machined with
the inverse groove pattern, and used to collect copper ions in a
low current bath. This technique guarantees very high precisions. The 
inner corrugation profile has the dimensional tolerances of the master as 
the electroformed material follows its profile at molecular level. 
The master is eventually removed with a corrosive chemical solution,
and the corrugated internal surface of antenna is cleaned by cycling
it in an ultra-sound device. Finally the feed horn is gold plated. 
Figure \ref{fig:mandrino} reports the difference between the measured and theoretical 
100 GHz mandrel dimensions as function of the corrugation number from the throat to the aperture: an accuracy $<10\mu$m can be reached.

As the 30 GHz feed horn has larger dimensions, the electro--erosion process of an 
aluminum cylinder has been used to manufacture it. The choice of aluminum electro--erosion has mainly the advantage of 
reducing the feed horn mass, which is very important for their use on board a satellite. 
Electro--erosion, also known as electrical discharge machining, is
a process by which conductive materials can be
removed from a metal by an electric spark. The 30 GHz
prototype has been manufactured by electro-erosion of a piece of
Aluminum alloy 6061, frequently employed for cryogenic
space applications.

%%%%%%%%%%%%%%%%%%%%%%
\section{Measurements}
%%%%%%%%%%%%%%%%%%%%%%
The measurements were performed with a Millimetre Vector Network
Analyzer\footnote{AB--millimetre Model MVNA 8-350; {\tt http://www.abmillimetre.com}}
(MVNA) in a semi-anechoic environment, obtained using a
screen of $^\copyright$Eccosorb panels around a wooden bench where
the test was made. A standard conical horn, with a directivity
nearly 21 dBi, was placed at about one meter from the aperture of
the feed horn under test and was used as the fixed receiver.
With the adopted setup, standing wave effects have been checked to have a 
negligible effects on the pattern measurements.

The feed horn was mounted on a rotation stage connected to the MVNA,
that can drive the angular sweeps directly. Two VNA heads
measured the transmission between the horn under test and the 
measuring one at fixed frequency.

Two translation stages were mounted above the rotation stage. They 
allowed the feed horn to be moved with good
precision along the x,y coordinates in the plane perpendicular to the 
rotation axis.
The x movement (perpendicular to the horn axis) was required for 
alignment, to adjust the position of the horn so that its axis was 
exactly intersecting the rotation axis. Otherwise the phasefront measurements 
would be biased by a spurious contribution due to misalignment.
The degree of freedom along the axis of the horn (i.e. y) was used to evaluate the position 
of the phase center from the measured phase pattern, which was exactly flat when 
the rotation axis and the phase center were coincident. 

Even if the translation stages
have 0.01 mm precision, the overall uncertainty is given by the
positioning of the horn on the mounts, therefore only $\pm$ 0.5
mm.

\subsection{Beam Pattern}

\begin{figure}[!ht]
\centerline{\includegraphics[width=28pc]{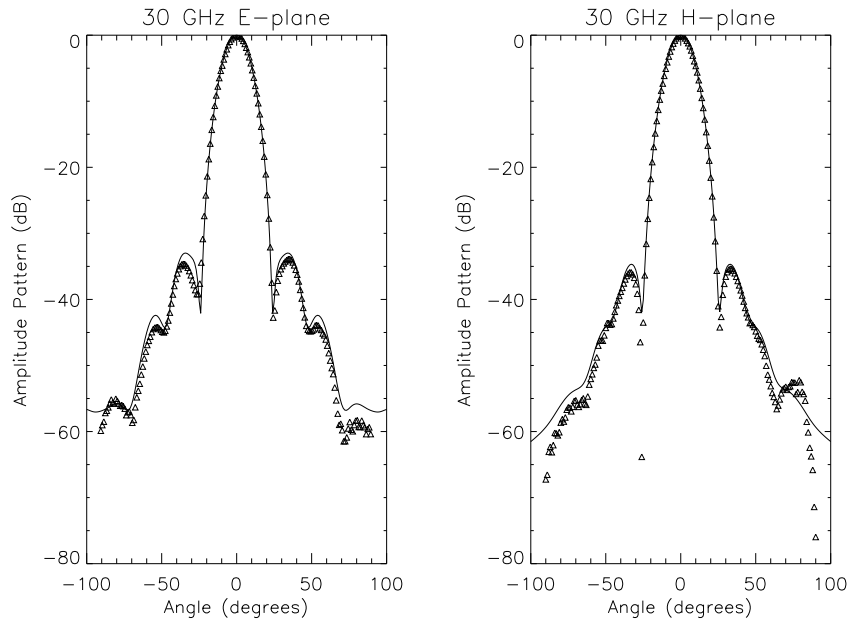}}
\caption{Comparison between E and H plane patterns measured (triangles) and simulated (solid line) at 30 GHz.}
\label{fig:30GHz}
\end{figure}

\begin{figure}[!h]
\centerline{\includegraphics[width=28pc]{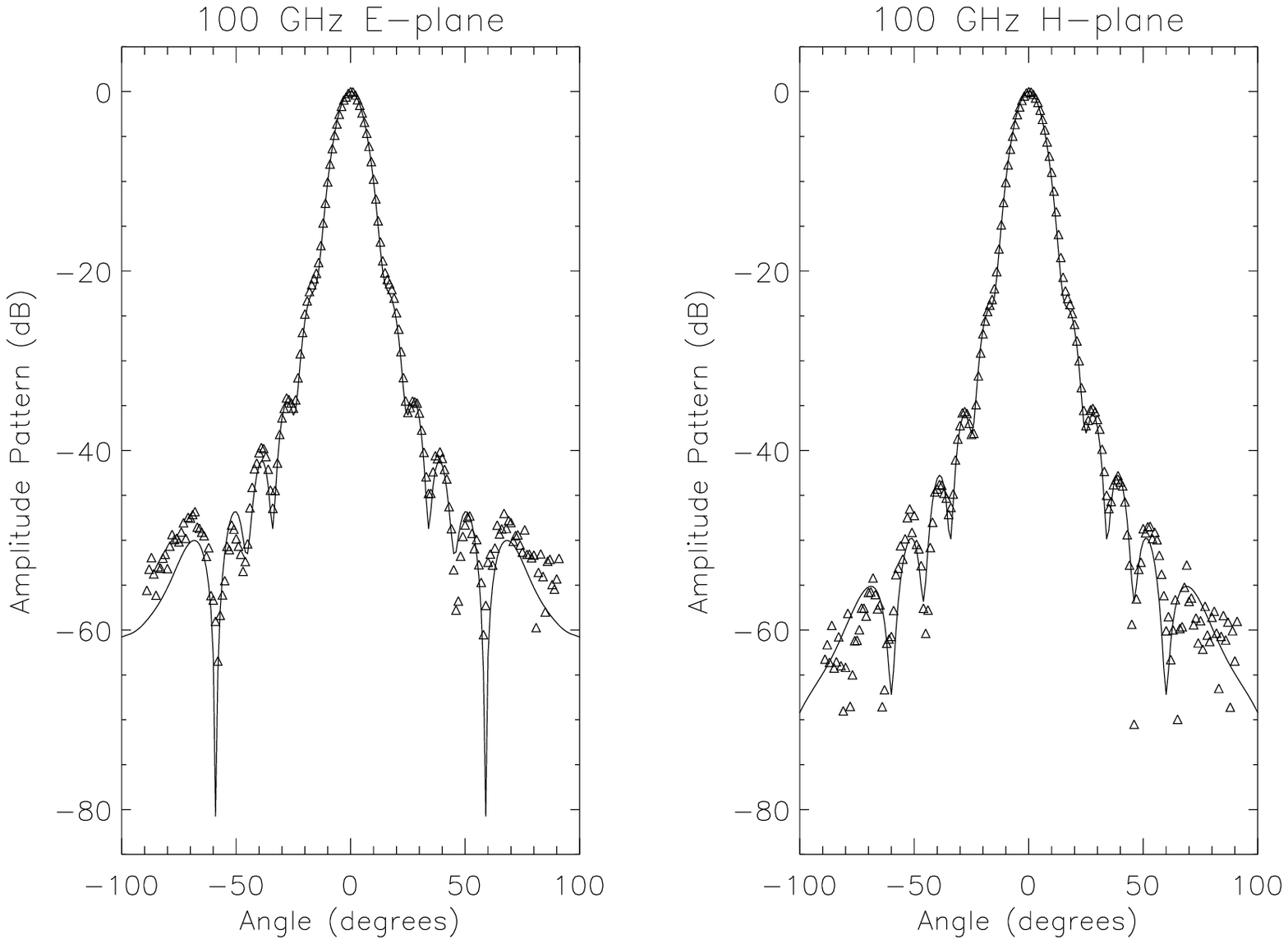}}
\caption{Comparison between E and H plane patterns measured (triangles) and simulated (solid line) at 100 GHz.}
\label{fig:100GHz}
\end{figure}

The radiation pattern in the principal (E and H) planes for the
30, 70 and 100 GHz EBB feed horns are shown in Figures
\ref{fig:30GHz} $\div$ \ref{fig:100GHz}. Typical measurement uncertainty ranges from $\pm 0.2$ dB to $\pm 0.5$ dB as the power decreases.
A comparison with the electromagnetic simulations based on Mode Matching (MM) techniques
combined with a Moment Method (MoM) code is also reported
for each plane. 
The far sidelobe level (at
off-axis angles greater than $80$ degree) is about $-60$ dB for the
feed horns at $30$ and $70$ GHz, while it is about $-55$ dB for the feed
horn at $100$ GHz. The expected sidelobe levels have been met in all
cases. It should be noted that the simulations do not take 
into account the real geometry of the setup like the $^\copyright$Eccosorb surrounding the horn aperture and supporting structures. Thus, differences between the simulations and the measurements may arise in the sidelobes as noticed in Figure \ref{fig:30GHz}.

The main lobe of each beam pattern shows a high degree of
symmetry, as expected. A very good symmetry is achieved also at large offset angles, for each feed. 

\begin{figure}[!h]
\centerline{\includegraphics[width=28pc]{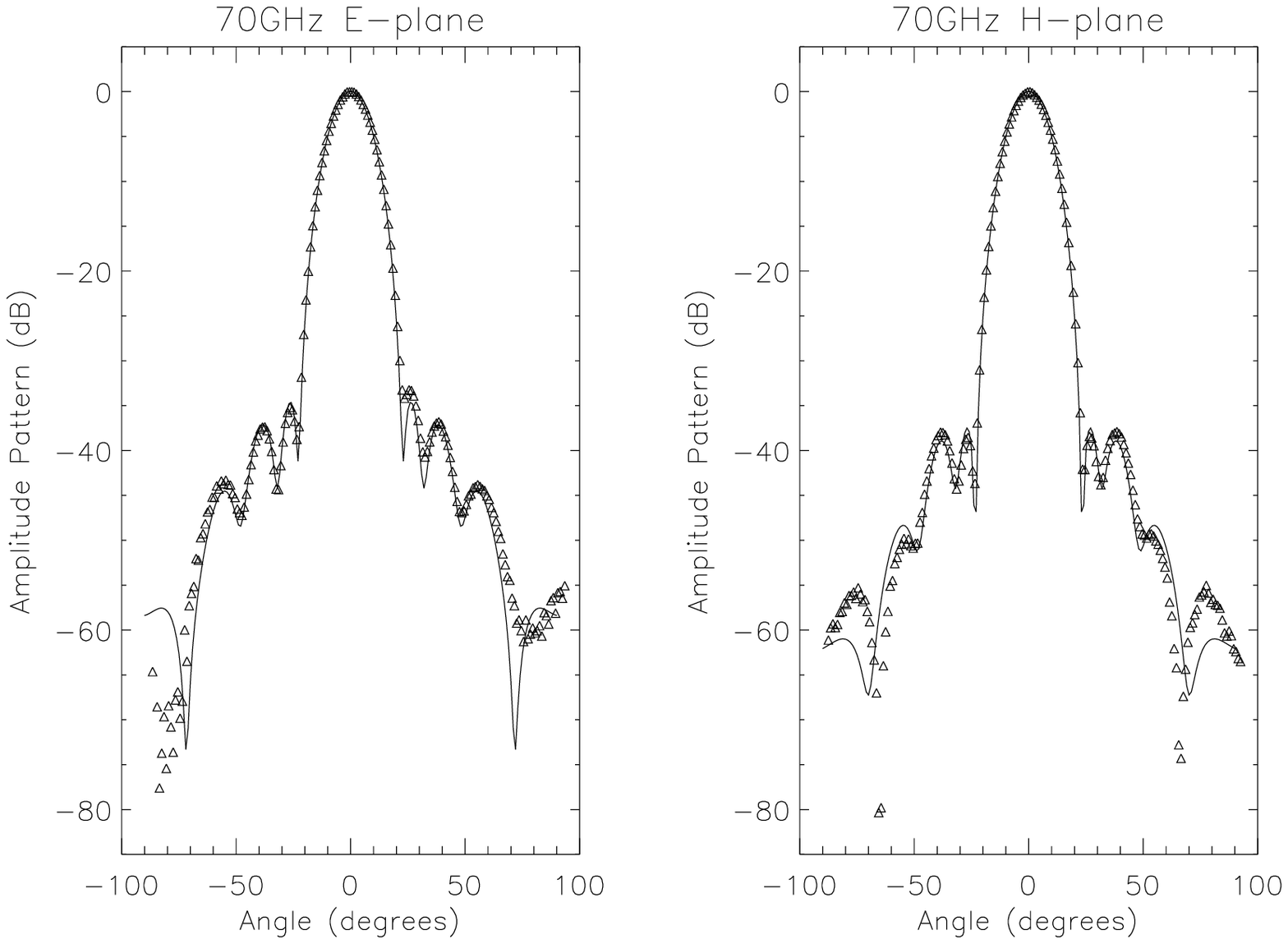}}
\caption{Comparison between E and H plane patterns measured (triangles) and simulated (solid line) at 70 GHz.}
\label{fig:70GHz}
\end{figure}

\subsection{Return Loss}
\begin{figure}[!ht]
\centerline{\includegraphics[width=28pc]{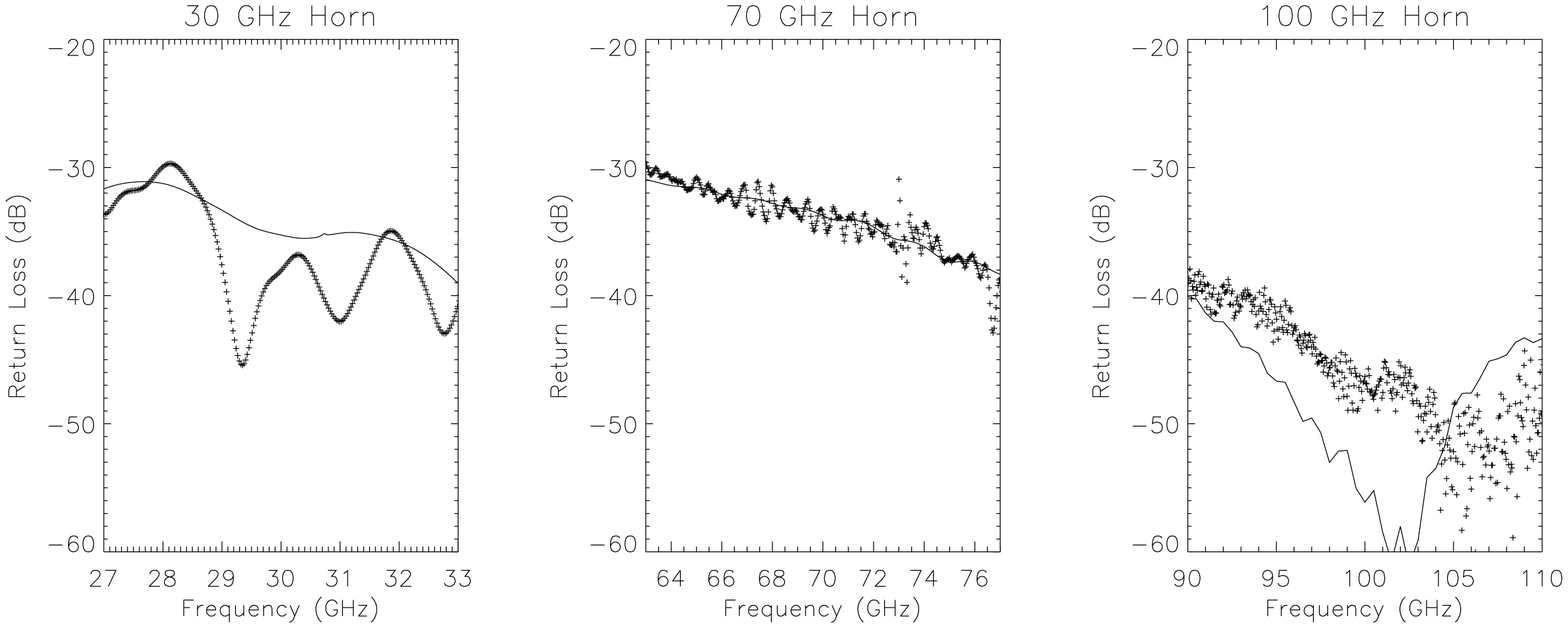}} \caption{Return
loss of the EBB horns; measured data (plus signs) and
simulations(solid line). Left: 30 GHz, middle 70 GHz and right 100
GHz.} 
\label{fig:RL}
\end{figure}

Return loss measurements were made using a non standard
rectangular-circular transition connecting the horn to the MVNA. 
Time-domain analysis was used to remove the effects of external reflections and the
transition. A wide frequency sweep was Fourier transformed, and the resulting time series was cleaned in regions corresponding (via the group velocity) to the transition 
and environmental contributions. In the design phase of the horns, a return loss better than $-30$dB 
over the full bandwidth has been required. Although a $5$dB of return loss degradation due to 
imperfections and flange connections is still tolerated, Figure \ref{fig:RL} shows that the measured return loss of the three horns is close to the predictions and then well within requirements. 

\subsection{Phase Center}

Phase patterns in the principal (E and H) planes have been taken
for several displacements of the EBB feed horn under test along
its axis, toward the receiver. The curvature of the phase front
was estimated with a parabolic fit of its central region, where
the pattern can be assumed Gaussian. The coefficient of the
quadratic term was plotted versus the displacement of the rotation
axis from the aperture. The zero-crossing position marks the point
where the rotation axis is exactly through the phase center. 
The phase centres of the three horns have been measured in this way.

For the $30$ GHz feed horn the phase centre resulted at $-10.7\pm 0.5$ mm and $-13.3\pm 0.5$ mm from the aperture plane for the E plane and H plane, respectively.
For comparison, the requirement has been assumed to be  $-12.39$ mm below the aperture plane.

For the 70 GHz horn, the phase center location resulted at $-2.11\pm 0.5$ mm for the E plane and $-0.71\pm 0.5$ mm for the H plane. The requirement of the phase centre has been assumed to be at $-2.36$ mm.

At 100 GHz, the phase centre resulted at $-9.87$ mm for the E plane and $-6.33$ mm for the H plane. For this horn the requirement has not been specified.

%%%%%%%%%%%%%%%%%%%%
\section{Conclusion}
%%%%%%%%%%%%%%%%%%%%

Three high performance corrugated conical feed horns at 30 GHz, 70
GHz and 100 GHz have been designed, fabricated, and tested, as
advanced prototypes for the Low Frequency Instrument (LFI) of the
ESA {\sc Planck} mission. The horns include linear and dual--profiled 
designs, and both electro--forming and electro--erosion fabrication techniques. 

The 100 GHz horn has been designed as a
linear corrugated feed horn while 30 and 70 GHz were dual profiled
corrugated horns. The 30 GHz prototype has been manufactured by
electro-erosion and the fabrication at higher frequencies has been
achieved by electro-formation. The position of the phase center
has been evaluated for each feed horn. The measured performances
on side-lobes and return loss are met over the large (20\%)
bandwidth required by the LFI high sensitivity measurements.
All the prototypes showed a return loss well below $-30$dB 
($-25$dB if the degradation due to the flanges is considered), as required. 
The side lobe level, as expected, resulted not exceeding 
the $-35$dB below the peak. Accurate horn models are required in order to compute the optical response of the entire feed--telescope system. Although the cross--polar field of these prototypes has not been measured, the simulation results indicate that the maximum level of cross polarization does not exceed -30 dB. However, in the case of {\sc Planck}/LFI, the cross--polar level is dominated by the telescope, since the LFI feeds are located far from the optical axis where aberrations become important. 

Typically, the predicted pattern (in amplitude and phase) of the horns are used as input for reflector antenna codes in order to simulate the radiation pattern of the telescope. These codes require that the horn pattern must be known in several azimuthal cuts, both in amplitude and phase, in order to accurately predict the currents on the reflectors. This is not easy to do by measurements. The excellent agreement between the simulated and measured pattern, as in the case of the horns described here, 
allows the use of the simulated pattern for precise optical calculations. 

\begin{acknowledgements}
We wish to thanks Prof. G. Pelosi - Dipartimento di Elettronica e Telecomunicazioni Universit\`a di Firenze and Prof. G. Tofani - IRA-CNR for the CAD facilities they made available to support the development of the dual profile horn design. 
\end{acknowledgements}

\end{article}

\begin{thebibliography}{}

\bibitem[\protect\citeauthoryear{Barnes et al.}{2002}]{Barnes02}

Barnes, C., Limon, M., Page, L., Bennett, C.L., Bradley, S., Halpern, M., Hinshaw, G., 
Jarosik, N., Jones, W., Kogut, A., Meyer, S., Motrunich, O., Tucker, G., Wilkinson, D., 
Wollack, E.:
2002,'The MAP Satellite Feed Horns', {\it ApJS}, {\bf Vol.143}, pp. 567--576.

\bibitem[\protect\citeauthoryear{Bennett et al.}{2003}]{Bennett03}
Bennett,C.L., Bay, M., Halpern, M., Hinshaw, G., Jackson, C., Jarosik, N., Kogut, A., 
Limon, M., Meyer, S.S., Page, L., Spergel, D.N., Tucker, G.S., Wilkinson, D.T., Wollack, E., Wright, E.L.: 2003, 'The Microwave Anisotropy Probe Mission',{\it ApJ}, {\bf Vol.583}, pp. 1--23.

\bibitem[\protect\citeauthoryear{Bersanelli et al.}{1998}]{Bersanelli98}
Bersanelli, M., Mattaini, E., Santambrogio, E., Simonetto, A.,
Cirant, S., Gandini, F., Sozzi, C., Mandolesi, N., Villa, F.:
1998, 'A low-sidelobe, high frequency corrugated feed horn for CMB
observations',{\it Experimental Astronomy}, {\bf Vol.8}, pp. 231--238.

\bibitem[\protect\citeauthoryear{Clarricoats and Olver}{1984}]{Clarricoats84}
Clarricoats, P.J.B. and Olver, A.D.: 1984, 'Corrugated horns for
microwave antennas', London.

\bibitem[\protect\citeauthoryear{Coccioli et al.}{1996}]{Coccioli96}
Coccioli, R., Pelosi, G., Ravanelli, R.: 
1996, 'Combined mode matching-integral equation technique for
feeders optimization',{\it in Software for Electrical Engineering Analysis and Design. Editor P.P. Silvester}, Southampton (UK): Computational Mechanics Publications.

\bibitem[\protect\citeauthoryear{Dubruel et al.}{2000}]{Dubruel00}
Dubruel, D., Cornut, M., Fargant, G., Passvogel, T., De Maagt, P.,
Anderegg, M., Tauber, J.: 2000, 'Very Wide Band Antenna Design for
Planck Telescope Project Using Optical and Radio Frequency
Techniques', {\it Millennium Conference on Antennas \& Propgation
- AP2000}, ESA SP-444 Proceedings.

\bibitem[\protect\citeauthoryear{Gentili et al.}{2000}]{Gentili00}
Gentili, G.G., Nesti, R., Pelosi, G., and Natale, V.: 2000,
'Compact dual-profile corrugated circular waveguide horn', {\it
Electronics Letters}, {\bf Vol.36, No.6.}, pp. 486--487.

\bibitem[\protect\citeauthoryear{Guzzi et al.}{1999}]{Guzzi99}
Guzzi, P.; Bersanelli, M.; Mattaini, E.; Santambrogio, E.;
Gandini, F.; Muzzini, V.; Simonetto, A.; Sozzi, C.; Villa, F.;
Mandolesi, N.: 1999, 'Fabrication and testing of W-band Corrugated
Feedhorns For CMB Measurements',{\it ITESRE/CNR Internal
Report},{\bf 237}.

\bibitem[\protect\citeauthoryear{Mandolesi et al.}{1998}]{Mandolesi98}
Mandolesi, N., Bersanelli, M.: 1998, 'The Low Frequency
Instrument', {\it Announcement of Opportunity for the FIRST/Planck
Programme}.

\bibitem[\protect\citeauthoryear{Olver and Xiang}{1988}]{Olver88}
Olver, A.D., and Xiang, J.: 1988, 'Design of Profiled Corrugated
Horns', {\it IEEE Trans. On Antenna Propagation}, {\bf Vol.36,
No.7.}, pp. 936--940.

\bibitem[\protect\citeauthoryear{Olver}{1992}]{Olver92}
Olver, A.D.: 1992, 'Corrugated horns',{\it Electronics \&
Communication Engineering Journal}, pp. 4--10.

\bibitem[\protect\citeauthoryear{Olver et al.}{1994}]{Olver94}
Olver, A.D., Clarricoats, P.J.B., Kishk, A.A, Shafai, L.: 1994, 
'Microwave Horns and Feeds',{\it IEE Electromagnetic Waves Series 39}

\bibitem[\protect\citeauthoryear{Puget et al.}{1998}]{Puget98}
Puget, J-L, Lamarre, J-M.: 1998, 'The High Frequency Instrument',
{\it Announcement of Opportunity for the FIRST/Planck Programme}.

\bibitem[\protect\citeauthoryear{Sletten}{1988}]{Sletten88}
Sletten, C.J.: 1988, 'Reflector and Lens Antennas, Analysis and Design Using Personal Computers', {\it Artech House Publishers}, pp. 78--84.

\bibitem[\protect\citeauthoryear{Teniente-Vallinas et al.}{2001}]{Teniente01}
Teniente-Vallinas, J.; Gonzalo-Garcia, R.; del-Rio-Bocio, C.: 2001,
'Ultra-wide band corrugated Gaussian profiled horn antenna design', 
{\it Antennas and Propagation Society, 2001 IEEE International Sym}, {\bf Vol.2},
pp. 316--319.

\bibitem[\protect\citeauthoryear{Toral et al.}{1989}]{Toral89}
Toral, M.A.; Ratliff, R.B.; Lecha, M.C.; Maruschak, J.G.; Bennett,
C.L.; Smoot, G.F.: 1989, 'Measurements of very low-sidelobe
conical horn antennas',{\it IEEE Trans. On Antennas and
Prop.},{\bf Vol.37, No.2.} pp. 171--177.

\bibitem[\protect\citeauthoryear{Villa et al.}{1997}]{Villa97}
Villa, F.; Bersanelli, M.; Mandolesi, N.: 1998, 'Design of Ka and
Q Band Corrugated Feed Horns For CMB Observations',{\it ITESRE/CNR
Internal Report},{\bf 188}.

\bibitem[\protect\citeauthoryear{Villa et al.}{1998}]{Villa98}
Villa, F.; Bersanelli, M.; Mandolesi, N.: 1998, 'Design of a
W-band Corrugated Feed Horn For The Beast Telescope',{\it
ITESRE/CNR Internal Report},{\bf 206}.

\bibitem[\protect\citeauthoryear{Villa et al.}{2001}]{Villa01}
Villa, F., Bersanelli, M., Burigana C., Butler, R.C., Mandolesi, N., et al.: 
'The Planck Telescope',{\it Proceedings of the 2K1BC Workshop}, 
AIP Conference Proceedings {\bf Vol. 616}, pp. 224--228. 

\bibitem[\protect\citeauthoryear{Villa et al.}{2003}]{Villa03}
Villa, F., Mandolesi, N., Butler, R.C.: 
'The Planck Project',{\it Mem S.A.it.}, {\bf Vol.74}, pp. 223--229 


\bibitem[\protect\citeauthoryear{Xiaolei}{1993}]{Xiaolei93}
Xiaolei, Z.: 1993, 'Design of Conical Corrugated Feed Horns for Wide-Band High-Frequency
Applications',{\it IEEE Trans. On Microwave Theory and Techniques},{\bf Vol.41, No.8.}
pp. 1263--1274.


\end{thebibliography}
\end{document}